\documentclass[aps,prd,preprint,superscriptaddress]{revtex4-1}  
\usepackage{latexsym,amsmath,amssymb,amstext}
\usepackage{slashed}
\usepackage{epsfig}
\usepackage{float,xcolor}
\usepackage{graphicx}
\usepackage{dcolumn}

\newcommand{\be}{\begin{equation}}
\newcommand{\ee}{\end{equation}}
\newcommand{\bea}{\begin{eqnarray}}
\newcommand{\eea}{\end{eqnarray}}
\newcommand\bef{\begin{figure}}
\newcommand\eef[1]{\label{fg:#1}\end{figure}}
\newcommand\beq{\begin{equation}}
\newcommand\eeq[1]{\label{#1}\end{equation}}
\newcommand\beqa{\begin{eqnarray}}
\newcommand\eeqa[1]{\label{#1}\end{eqnarray}}
\newcommand\bet{\begin{table}}
\newcommand\eet[1]{\label{tb:#1}\end{table}}

\newcommand\fgn[1]{Figure \ref{fg:#1}}
\newcommand\eqn[1]{Eq.\ (\ref{#1})}

\newcommand\tbn[1]{Table \ref{tb:#1}}

\newcommand\ie{{\sl i.e.\/}}

\begin{document}

\widetext

\title{Parity anomaly cancellation in a three-dimensional QED with single massless Dirac fermion}

\author{Nikhil\ \surname{Karthik}}
\email{nkarthik@bnl.gov}
\affiliation{Physics Department, Brookhaven National Laboratory, Upton, New York 11973-5000, USA}
\author{Rajamani\ \surname{Narayanan}}
\email{rajamani.narayanan@fiu.edu}
\affiliation{Department of Physics, Florida International University, Miami, FL 33199}

\begin{abstract}
We study a three-dimensional non-compact QED with a single two-component
massless fermion and two infinitely massive regulator fermions of
half the charge using lattice overlap formalism. The parity anomaly
is expected to cancel exactly between the massless and regulator
fermions in the continuum, but this cancellation is inexact on
lattice akin to lattice chiral gauge theories.  We show non-perturbatively
that parity-breaking terms vanish in the continuum limit at any
finite volume. We present numerical evidences that the resulting
parity-invariant theory spontaneously breaks parity in the infinite
volume limit.
\end{abstract}

\date{\today}
\maketitle

\section{Introduction}
The standard model of particle physics is anomaly free due to an
exact non-trivial cancellation of gauge anomalies~\cite{Schwartz:2013pla}
from different representations to all orders of perturbation theory.
Chiral anomalies outside perturbation theory can be discussed
geometrically~\cite{AlvarezGaume:1983cs} and the relation between
consistent and covariant currents~\cite{Bardeen:1984pm} plays a
central role.  Such fundamental issues should be addressed in any
non-perturbative formalism of chiral gauge theories. Overlap formalism
of chiral gauge theories on the lattice~\cite{Narayanan:1994gw} was
motivated~\cite{Narayanan:1992wx} by an attempt to regularize a
specific chiral gauge theory using infinite number of Pauli-Villars
fields~\cite{Frolov:1992ck} and the ability to use domain walls to
create a chiral zero mode~\cite{Kaplan:1992bt}. In order to discuss
the problem of chiral anomalies in a gauge covariant and geometric
manner, a two-form in the space of gauge fields defined through the
curl of the difference between the covariant and consistent currents
was introduced within the overlap formalism in~\cite{Neuberger:1998xn},
and it was identified to be the Berry's curvature.  Two sources
contribute to this Berry's curvature for a chiral fermion in an
anomalous representation -- the first is due to the genuine continuum
gauge anomaly that cannot be removed, and the second is due to the
spatial smearing of the anomalous contribution due to finite lattice
spacing. There is just the  contribution due to smearing in an
anomaly free chiral theory which can only be removed by fine-tuning
the irrelevant terms in fermion action on the
lattice~\cite{Neuberger:1998xn}. The exceptions to the fine-tuning
are QCD-like vector theories where the anomaly cancellation is
trivial.

The odd-dimensional analog to chiral anomalies is parity
anomaly~\cite{Deser:1981wh,Deser:1982vy,Niemi:1983rq,Redlich:1983dv} and
this also can be discussed geometrically~\cite{AlvarezGaume:1984nf}.
In this letter, we consider a three-dimensional
analog to the chiral gauge theories, where there is a non-trivial
cancellation of parity anomaly between massless fermions and
infinitely massive fermions, which is a property unique to
three-dimensions.  The theory we consider is an Abelian $U(1)$ gauge
theory with one massless Dirac fermion of charge $q$ and two
infinitely massive fermions of charges $\frac{q}{2}$ in a three-torus
with physical size, $\ell^3$. This corresponds to the Euclidean
continuum theory, with an implicit regularization,
\be
{\cal L}=\overline{\psi}\left(\slashed{\partial}+iq
\slashed{A}\right)\psi-\frac{q^2 i}{8\pi}\epsilon_{\mu\nu\rho}A_\mu\partial_\nu A_\rho+\frac{1}{4}F_{\mu\nu}F^{\mu\nu},
\label{model}
\ee
written in standard notation in units where the coupling constant
$g^2=1$.  This theory has phenomenological relevance to the low-energy
physics of fractional quantum Hall effect at half-filled Landau
level~\cite{Son:2015xqa,Metlitski:2015eka,Wang:2016fql}.  Like in
even dimensions, lattice regularization of this theory within the
overlap formalism~\cite{Narayanan:1994gw,Kikukawa:1997qh,Narayanan:1997by}
does not succeed in an exact cancellation of the parity anomaly.
A salient result in this letter is the numerical evidence for the
restoration of parity invariance in the continuum at any finite
physical volume without the need for fine-tuning the fermion action,
which suggests a similar situation to hold in even dimensional
chiral gauge theories as anticipated in~\cite{Neuberger:1998xn}.
This will also establish the existence of such three-dimensional
theories outside perturbation theory.  We will then present a
numerical study of this theory in the infinite volume limit and
provide evidence for spontaneous breaking of parity.

\section{Modus operandi} 
As is standard in lattice field theory, we discretize the physical
volume $\ell^3$ using $L^3$ lattice points with the lattice spacing
being $\frac{\ell}{L}$.  The continuum limit is achieved by taking
the $L\to\infty$ limit at fixed value of $\ell$.  For the Abelian
theory, the dynamical real lattice variables are $\theta_\mu({\bf
n})$ at the link connecting the lattice point at $\mathbf{n}$ to
$\mathbf{n+\hat\mu}$.  The lattice regularized partition function
of the model in \eqn{model} using the overlap
formalism~\cite{Kikukawa:1997qh,Narayanan:1997by} is
\be
Z(\ell,L) 
= \int [d\theta] e^{-S_g(\theta)}  \det\left(\frac{1 + V_{\theta}}{2}\right) {\det}^2 V^\dagger_{\frac{1}{2}\theta},
 \label{sonmodel}
\ee
where $S_g(\theta)$ is the non-compact gauge action on the lattice
(obtained by discretizing the $F^2_{\mu\nu}$ term).  The unitary
operator $V_{q\theta}$ depends on the compact link variables
$U^q_\mu(\mathbf{n})=e^{iq\theta_\mu({\bf n})}$ where $q$ is the charge
of the fermion coupled to the gauge field. We have set $q=1$ in
\eqn{model} and the first determinant factor realizes the effective
action obtained by integrating out the massless fermion in \eqn{model}
and the second determinant factor realizes the Chern-Simons term
in \eqn{model} as induced by an infinitely massive fermion.

If we define the induced action $2{\cal A}_q$ from the infinite
mass fermion via, $\det V_{q\theta}\equiv \exp\left(2 i q^2 {\cal
A}_q\right)$, then we expect ${\cal A}_q(\theta)$ to be independent
of $q$ for smooth gauge
fields~\cite{Redlich:1983dv,Niemi:1983rq,Coste:1989wf,Karthik:2015sza}, and
be the same as the level-one Chern-Simons action.  If we perform
the Euclidean parity transformation, under which $V_{q\theta}\to
V^\dagger_{q\theta}$, the path integral in \eqn{sonmodel} transforms
to
\be
Z(\ell,L) =   
\int [d\theta] e^{-S_g(\theta)}  \det \frac{1 + V_{\theta}}{2} {\det}^2 V^\dagger_{\frac{1}{2}\theta} e^{-2 i{\cal A}(\theta)},
\label{zmp}
\ee
where
\be
{\cal A}(\theta)={\cal A}_1(\theta)-{\cal A}_{\frac{1}{2}}(\theta).
\label{cala}
\ee
Parity anomaly cancellation in the continuum means that
 $2{\cal A}=0$ or
equivalently, ${\cal A}=n\pi$ for $n=0,\pm 1$ as $L\to\infty$.  On
the lattice however, the non-trivial anomaly cancellation between
two different charges will result in $2{\cal A}(\theta)$ being zero
only on classically smooth backgrounds.  An ensemble of gauge field
configurations on the lattice away from the continuum limit will
not be smooth and we do not expect $2{\cal A}(\theta)=0\ ({\rm mod\
}2\pi)$, leading to
\be
 \det \frac{1 +  V_{\theta}}{2} {\det}^2 V^\dagger_{\frac{1}{2}\theta} = \left|\det \frac{1 + V_{\theta}}{2} \right| e^{i{\cal A}(\theta)};\ {\cal A}\in (-\pi,\pi],
\ee
which forms the core of the problem addressed in this letter.

Our strategy can be summarized as follows.
Using the Rational Hybrid Monte Carlo (RHMC)~\cite{Kennedy:1998cu,Clark:2003na,Duane:1987de}, 
an algorithm based on molecular dynamics evolution, 
we numerically simulate the theory on the lattice using the positive
definite measure  
\be
p_+(\theta)=\left|\det \frac{1 + V_{\theta}}{2}
\right| e^{-S_g},\label{p+meas}
\ee
and consider the phase $e^{i{\cal A}}$ to be part of the observables.
Our first aim is to study the distribution of ${\cal A}$ generated
at a given $\ell$ and $L$ and show that the distribution has a
tendency to approach a delta function for all $\ell$ as we take
$L\to\infty$.  As the lattice spacing increases with $\ell$ in a
range of numerically feasible values of $L$, we can only provide
reasonable numerical evidence for parity anomaly cancellation over
a limited but wide range of $\ell$. Our second aim in this letter
is to assume that parity anomaly cancellation holds for all values
of $\ell$ and study the infrared physics of the model in \eqn{model}
using $p_+(\theta)$ as the measure.

\bef
\begin{center}
\includegraphics[scale=1.2]{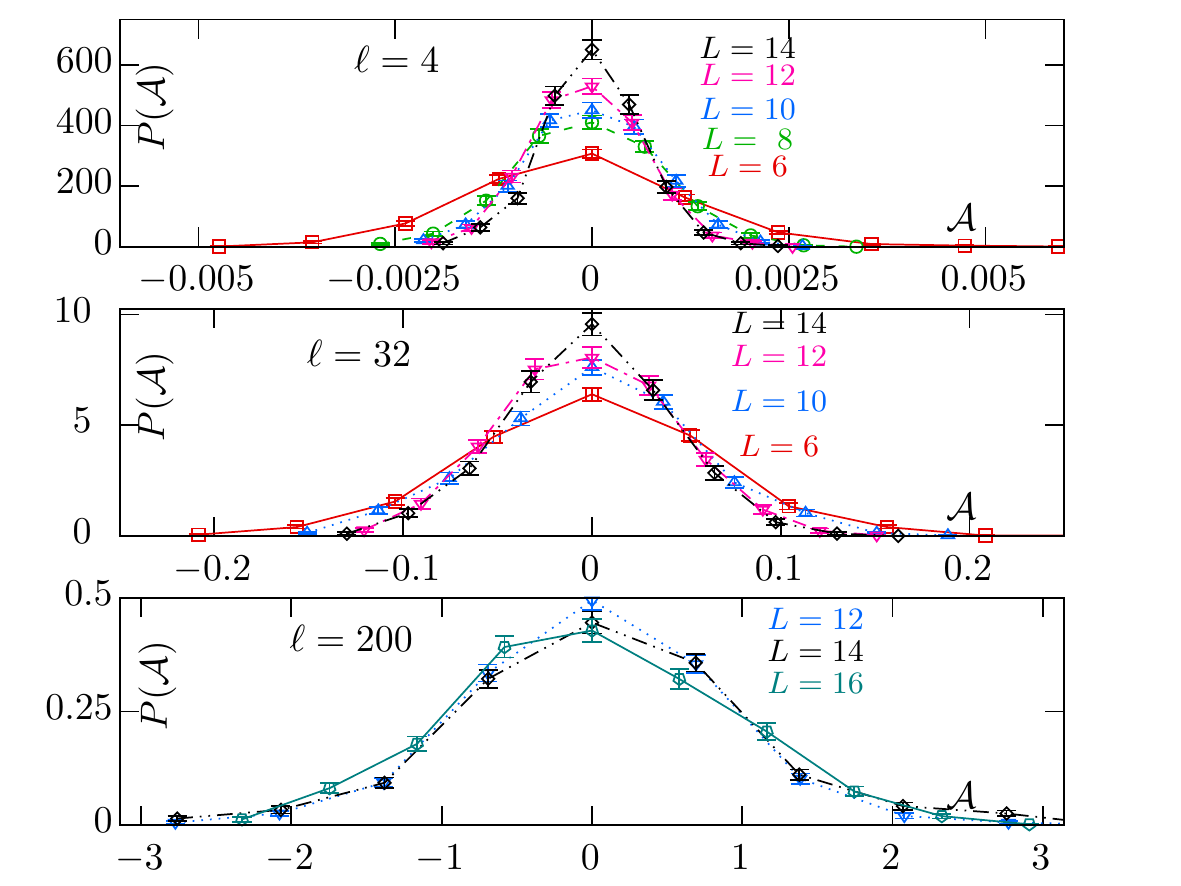}
\end{center}
\caption{The distribution of ${\cal A}(\theta)$ 
at different physical volumes $\ell^3$ are shown in the three panels. The different 
symbols correspond to different $L$.}
\eef{phase}

\section{Anomaly cancellation}
\fgn{phase}  shows the distribution $P({\cal A})$ of ${\cal A}(\theta)$
as sampled using $p_+(\theta)$ in three panels, top to bottom,  for
$\ell=4, 32, 200$ respectively.  Within each panel for a fixed
$\ell$, the different symbols correspond to different lattice
spacings.  Due to the parity-invariant measure $p_+(\theta)$, the
distributions are almost symmetric with small deviations resulting
from finite statistics.  We notice from the $\ell=4$ and $32$ panels
that $P({\cal A})$ gets sharper as one approaches the continuum
limit $L\to\infty$.  However, this approach of the width of the
distribution to zero is hard to see in the $\ell=200$ panel, and
it is understandable since the finest lattice spacing ($L=16$) at
$\ell=200$, where we were able to compute ${\cal A}$ is 5.4 times
larger than the one at $\ell=32$ ($L=14$).  By putting together the
data for ${\cal A}$ from all $\ell$ and $L$, we now justify that
the distributions at larger $\ell$ will indeed get sharper at
prohibitively large values of $L$.  Since one expects the remnant
phase ${\cal A}$ to be a volume integral of local irrelevant terms,
we show the variance per unit physical volume, $\ell^{-3}{\rm
Var}\left({\cal A}\right)$, as a function of lattice spacing,
$\ell/L$, in the left panel of \fgn{variance}.  The data points of
same colored symbol belong to a fixed value of $\ell$ but differ
in $L$, while different colored symbols correspond to different
$\ell$ as specified near them.  The data approximately falls on a
universal curve, with ${\rm Var}\left({\cal A}\right)\sim L^{-1}$ at smaller
$\ell/L$.   On the right panel of \fgn{variance}, we show the
scaled peak-height of the distribution, $\ell^{3/2}P({\cal A}=0)$,
as a function of $\ell/L$.  The approximate data collapse suggests
a $\sqrt{L}$ increase in the peak-height at smaller $\ell/L$. 
As expected, higher order effects in lattice spacing come into play
in both figures for larger $\frac{\ell}{L}$. 
Based on these empirical observations, we find reasonable evidence
for $P({\cal A})$ to approach a delta function in the continuum
limit at a fixed $\ell$ and it is important that one takes the
continuum limit before taking the infinite volume limit.

\bef
\begin{center}
\includegraphics[scale=1.2]{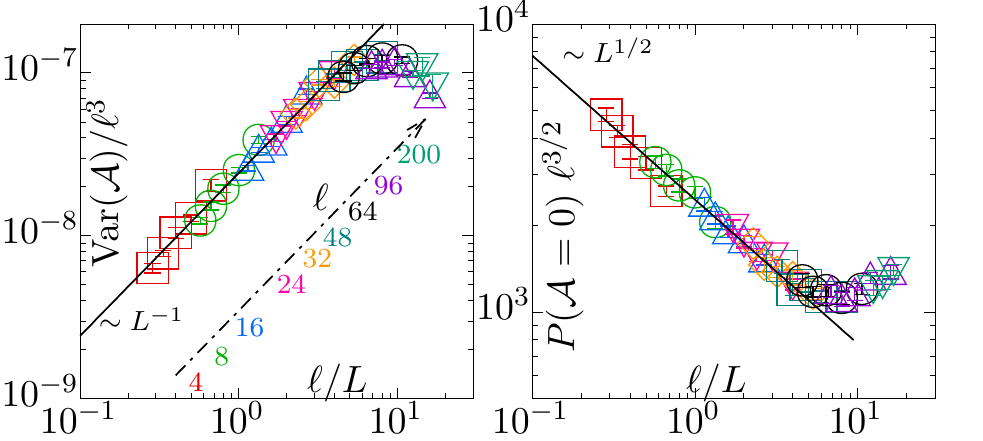}
\end{center}
\caption{The variance (left panel) and the height of the distribution
at ${\cal A}=0$ (right panel) for $P({\cal A})$, both scaled by
appropriate powers of $\ell$, are shown as functions of lattice
spacing.}
\eef{variance}

We now discuss the sign of the fermion determinant.  The distribution
$P({\cal A})$ on the coarser lattices, such as the one at $\ell=200$,
covers the entire range $(-\pi,\pi]$, but still remains peaked at
zero.  Based on the arguments above, this implies that the distribution
in the continuum limit will be peaked around zero, in spite of
values of ${\cal A}$ close to $\pi$ being allowed in the essentially
continuous molecular dynamics evolution of gauge fields used by the
RHMC algorithm on coarser lattices.  In principle, we could have
found a separation of our ensemble into two sectors on coarser
lattice spacings (corresponding to ${\cal A}(L=\infty)=0$ and
$\pm\pi$) easily identified by a doubly peaked $P({\cal A})$.  In
this case, it would have been necessary to have a zero of the fermion
determinant along the RHMC's canonical evolution as the continuum
limit is approached. Since we did not find this to be the case, our
result is consistent with the absence of topological zero modes in
odd-dimensional space without a
boundary~\cite{Atiyah:1963zz,Callias:1977kg}.  In this manner, we
have succeeded in demonstrating that  \eqn{sonmodel} has a parity
invariant as well as an effectively positive measure in the continuum.

\bef
\centering
\includegraphics[scale=1.2]{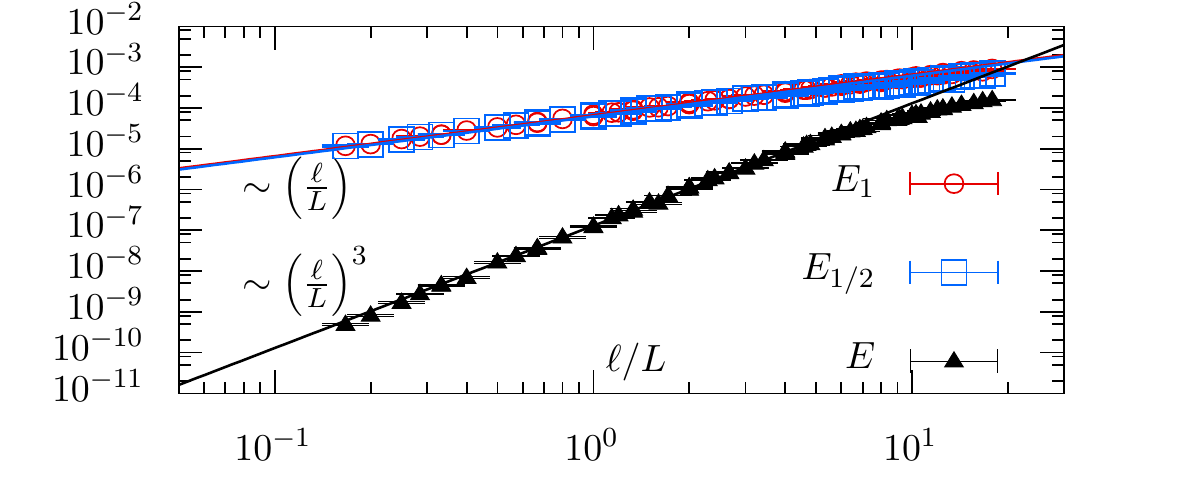}
\caption{The dependence of $E_1$,
$E_{1/2}$ and their difference, $E$,
on lattice spacing $\ell/L$.}
\eef{cscurrent}

Another quantity relevant to the anomaly cancellation is
\be
J^q_i(\mathbf{n})=\frac{\delta}{\delta\theta_i(\mathbf{n})} {\cal A}_q(\theta),
\label{current}
\ee
which is a fermion-induced pseudo-vector current in lattice units,
and the expectation value of its magnitude is $E_q({\bf n})=\langle
{\bf J}^q({\bf n}) \cdot {\bf J}^q({\bf n})\rangle_+$.  One expects
$J^q_i(\mathbf{n})$ to depend locally on the flux
$\sim\epsilon_{ijk}\Delta_j\theta_k$, but need not be ultra-local
and get smeared around ${\bf n}$ as discussed in~\cite{Neuberger:1998xn}.
In the absence of such an ultra-locality,
$E(\mathbf{n})=E_1(\mathbf{n})-E_{1/2}(\mathbf{n})$ will not vanish
at finite lattice spacing but it must vanish faster than $E_1(\mathbf{
n})$ and $E_{1/2}(\mathbf{n})$ as one approaches the continuum.  In
\fgn{cscurrent}, we put together the data from all $\ell$ and $L$
for $E_1$ and $E_{1/2}$ at an arbitrarily chosen $\mathbf{n}$, and
show it as a function of lattice spacing $\frac{\ell}{L}$.  The
data from different values of $\ell$ fall on the same curve due to
the local nature of this observable. The lattice spacing scaling
of $E_1$ and $E_{1/2}$ is $\frac{\ell}{L}$, the same as the average
local energy density.  With this combined data, we see that $E$
falls off with the lattice spacing like $\left(\frac{\ell}{L}\right)^3$,
faster than $E_1$ or $E_{1/2}$ by two powers of lattice spacing
ensuring again that the theory will be parity-invariant at all
values of $\ell$ studied here.

\bef
\begin{center}
\includegraphics[scale=1.2]{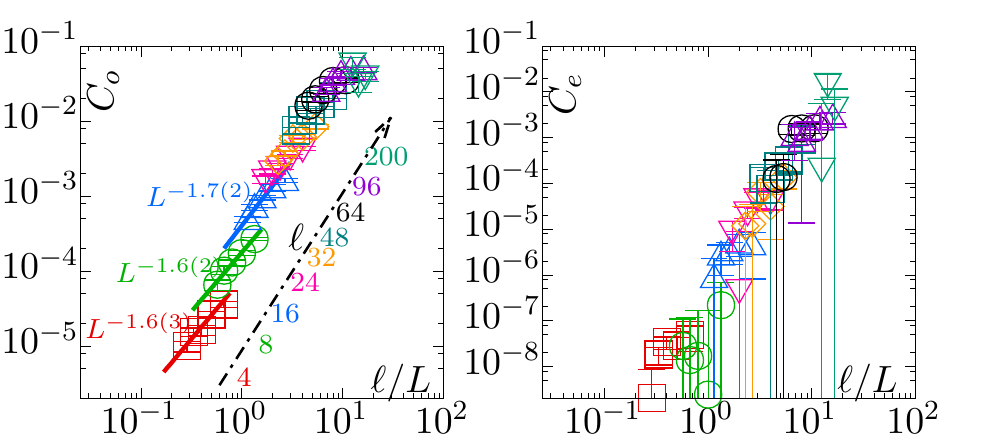}
\end{center}
\caption{The approach of $C_o$ (left) and $C_e$ (right) to 0 in the continuum limit.  
}
\eef{eplusminus}

Having demonstrated the path integral measure is anomaly free in the
continuum limit, it is also imperative that we show the VEVs of parity-odd
observables vanish in the continuum limit. Decomposing any observable
${\cal O}$ into its parity-even and odd components ${\cal O}_e$ and
${\cal O}_o$ respectively, its expectation value can be written as
\be
\langle {\cal O}(\theta) \rangle = \frac{\left\langle {\cal O}_e(\theta) \cos{\cal A}(\theta)\right\rangle_+}{\left\langle \cos {\cal A}(\theta) \right\rangle_+ } + i\frac{\left\langle {\cal O}_o(\theta) \sin{\cal A}(\theta)\right\rangle_+}{ \left\langle \cos {\cal A}(\theta) \right\rangle_+ }.
\ee
We want to show that in the continuum limit, the parity-even first
term on the right hand side becomes $\langle{\cal O}_e\rangle_+$
and the parity-odd second term vanishes. We consider the correction
$C_e + i C_o =\langle{\cal O}\rangle - \langle{\cal O}_e\rangle_+$
as a function of $L$.  For ${\cal O}$, we  used the dimensionless
lowest positive eigenvalue $\lambda^+_1(\theta) \ell$ of the inverse of
massless Hermitian overlap Dirac propagator,
$iG^{-1}(\theta)L=i\frac{1+V_{\theta}}{1-V_{\theta}}L$, at different
$L$.  In \fgn{eplusminus}, we show the decreasing
behavior of both $C_o$ and $C_e$ at different fixed $\ell$, as $L$
is increased.  The different colored symbols in the plot belong to
different $\ell$.  At any finite $L$, $C_o$ is significantly non-zero
and indeed decreases when the lattice spacing is made smaller. 
On finer lattices, a distinct $L^{-\Delta}$ behavior with 
an empirical value $\Delta\approx1.5$ is seen. For the data at larger $\ell/L$,
a downward curvature is seen implying the asymptotic values of 
$\Delta$ will be greater than what can be extracted from the data (which is about 1.2).
On other hand, the ultra-violet physics of
anomaly cancellation seems to decouple from the infra-red parity-even
expectation values, as seen from the fact that $C_e$ is much less
than $0.1\%$ of
$\left\langle\frac{1}{2}(\lambda^+_1(\theta)+\lambda^+_1(\theta_p))
\ell\right\rangle_+$ (and about two to three orders of magnitude
lesser than the corresponding $C_o$) in the range of $\ell$ we
studied.  In fact, for $L>10$, $C_e\approx 0$ within 1.5-$\sigma$
error range.

\bef
\begin{center}
\includegraphics[scale=1.2]{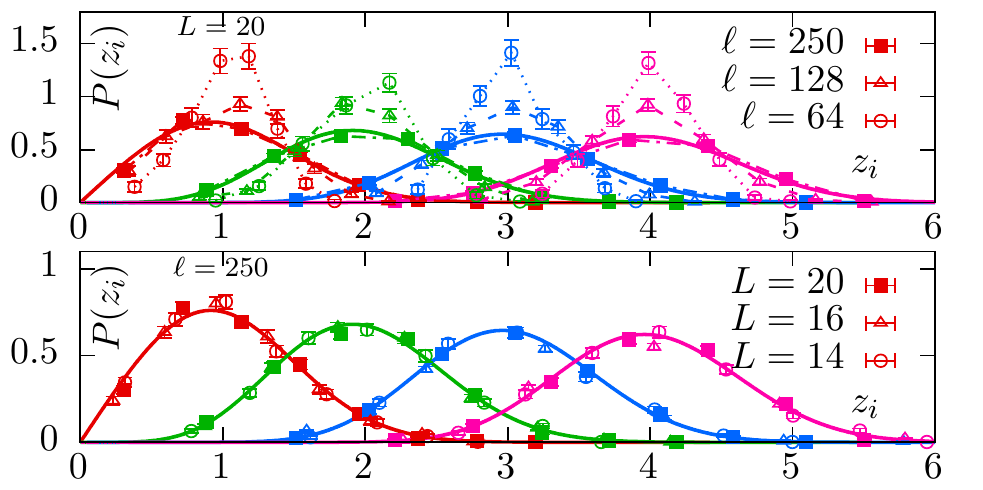}
\end{center}
\caption{
Comparison of the distributions of $\lambda_i\Sigma_i(\ell)\ell^3$
(symbols connected by dotted lines), and the RMT eigenvalues $z_i$
(solid curves).  The red, green, blue and purple curves correspond to $i=1,2,3,4$ respectively. 
Top: volume dependence at fixed $L=20$. Bottom:
lattice spacing dependence at $\ell=250$.
}
\eef{rmtdist}

\bef
\begin{center}
\includegraphics[scale=1.1]{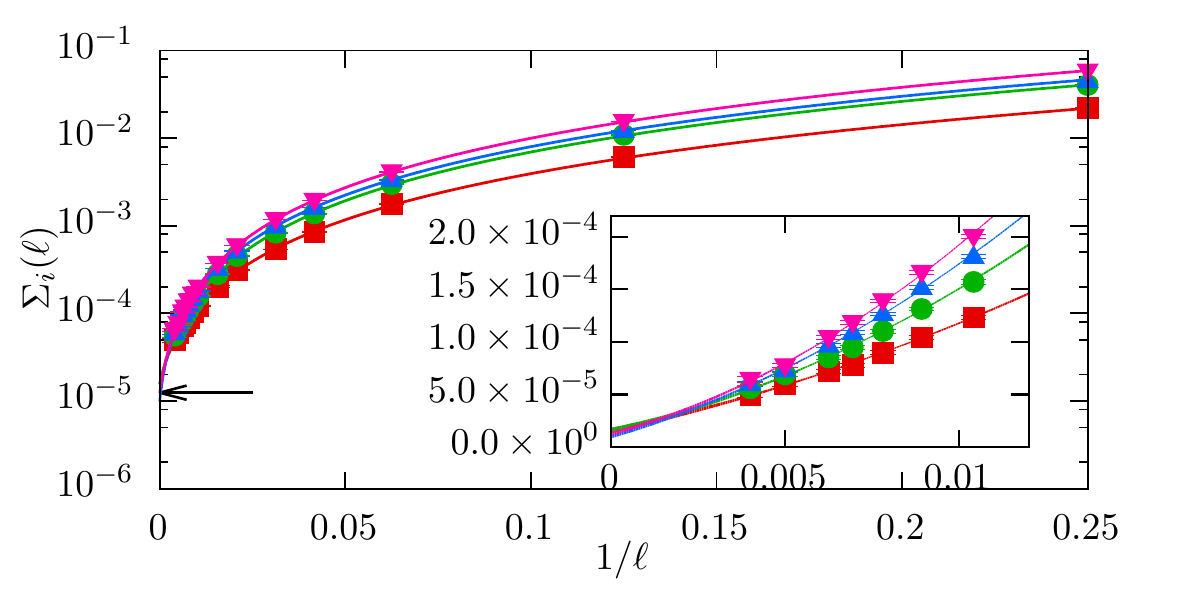}
\end{center}
\caption{
The infinite volume extrapolation of $\Sigma_i(\ell)$. 
The red, green, blue and purple points and curves correspond to $i=1,2,3,4$ respectively.
}
\eef{sigmaext}

\section{Spontaneous symmetry-breaking of parity}
Having numerically established a parity-invariant theory with a
positive measure in a certain range of $\ell$ that was numerically
accessible, we will assume this to be the case for higher values
of $\ell$ and study the infra-red behavior of the theory by taking
the $\ell\to\infty$ limit using the $p_+(\theta)$ measure. A
possibility is the spontaneous symmetry breaking (SSB) of parity
leading to a non-zero bilinear condensate $\Sigma$ \ie,  at finite
fermion mass $m$ and infinite volume, $\langle \bar\psi \psi
\rangle(m) = \Sigma\frac{m}{|m|}+O(m)$.  To study this, we focus
on the discrete dimensionless Dirac operator eigenvalues  ordered
by magnitude, $0< \ell\lambda_1(\theta;\ell) < \ell\lambda_2(\theta;\ell)
< \ldots,$ (which are technically obtained from
$L\left|G^{-1}(\theta)\right|$), at finite $\ell$. We first take
the $L\to\infty$ continuum limit of $\langle\lambda_i(\ell)\ell\rangle$
(using $L$ from 12 to 24) at different fixed $\ell$ ranging from 4
to 250 for this study before considering the $\ell\to\infty$ limit.

The probability distribution of $\lambda_i(\theta;\ell)$, as sampled in the 
Monte Carlo, will exhibit
several well separated peaks consistent with a spectrum that is
discrete.  Perturbation theory will hold as $\ell\to 0$ and
$\langle\lambda_i(\ell)\rangle$ will be proportional to $\ell^{-1}$.
If the theory spontaneously breaks parity as $\ell\to\infty$, then
$\langle\lambda_i(\ell)\rangle\sim\ell^{-3}$ (due to a finite
eigenvalue density near zero~\cite{Banks:1979yr}) and in addition,
the distributions of the individual eigenvalues should also match
with those from an appropriate random matrix theory (RMT)
ensemble~\cite{Verbaarschot:1994ip,Damgaard:1997pw,Nishigaki:2016nka}. 
If we define $\Sigma_i(\ell)$
through the means $\left\langle\lambda_i(\ell)\right\rangle$ and
$z_i$ of the two respective distributions,
\be
\left\langle\lambda_i(\ell)\right\rangle \Sigma_i(\ell) \ell^3 = z_i,
\label{sigdef}
\ee
then $\Sigma_i(\ell)$ for different $i$ should approach the same
non-zero value $\Sigma$ (the value of the condensate) as $\ell\to\infty$.

\fgn{rmtdist} shows a comparison of the distributions of the scaled,
four low-lying Dirac eigenvalues, $\ell^3\Sigma_i(\ell,L)\lambda_i(\theta;
\ell,L)$, to the distributions from the RMT, which are shown as
solid curves in the plots.  The top panel shows the volume dependence
of the distributions at a fixed number of lattice points $L=20$.
One can see the distributions approaching the RMT as $\ell$ is
increased from $\ell=64$ to $\ell=250$. The bottom panel shows this
agreement between the Dirac and RMT eigenvalue distributions at
$\ell=250$ is robust as the number of lattice points $L$ is made
larger from $L=14$ to 20.  A quantitative estimate shows that the
deviation of the data from the RMT distributions becomes smaller
with increasing $\ell$ and approaches zero in the infinite volume
limit. This agreement with RMT shows the presence of SSB.

\fgn{sigmaext} shows $\Sigma_i(\ell)$ as extracted from the matching
with RMT using \eqn{sigdef}, as a function of $\ell$.  The different
symbols are the values of $\Sigma_i(\ell); i=1,2,3,4$ in the continuum
at different fixed $\ell$. At any finite $\ell$ the values of
$\Sigma_i(\ell)$ from different $i$ do not agree, as expected.
Assuming the existence of finite non-zero value of the condensate
$\Sigma$ in the infinite volume limit, we used $\Sigma_i(\ell)=\Sigma_i+k_1
\ell^{-1}+k_2 \ell^{-2}$, to fit the entire range of finite $\ell$
data. These fits are shown by the curves.  The inset magnifies the
large $\ell$ region.  We find the extrapolated values of $\Sigma_i\times
10^{5}$ from $i=1,2,3,4$ to be $1.5(3), 1.5(3), 1.0(2)$ and $1.2(2)$
respectively.  Though the extrapolated values are about factor five
smaller than the available data point, the agreement between different
extrapolated values of $\Sigma_i$, together with the remarkable
agreement with RMT distributions are indications of a unique $\Sigma
\ne 0$ at infinite $\ell$.

\section{Discussion} 
An earlier attempt~\cite{Narayanan:1996cu} to verify cancellation
of anomalies in a two dimensional chiral gauge theory by directly
establishing gauge invariance in the continuum suffered from the
fact that there is no concept of smooth gauge transformations in
the continuum limit. In light of the results in this letter, it
would be interesting to revisit this problem by a computation of
the continuum limit of the Berry's curvature~\cite{Neuberger:1998xn}
in a sequence of lattice gauge field ensembles at different lattice
spacings.  Of experimental relevance are the response functions of
the single flavor theory studied here with the topological current
coupled to a background compact gauge field $\phi$, which can be
realized in our lattice setup by including the term $\det \left[
V_{\theta-\phi} V^\dagger_\theta V^\dagger_\phi \right]$ in
\eqn{sonmodel}.  This particular model appears in recent discussions
of duality between fermion
theories~\cite{Mross:2015idy,Cheng:2016pdn,Cordova:2017kue}.  It
would be interesting to see if the SSB has any effect on the induced
action for $\phi$.  It is trivial to extend the overlap formalism
presented here for three-dimensional QED with arbitrary number of
flavors ($N$) of massless Dirac fermions and arbitrary number of
flavors $(k)$ of infinite mass fermions.  This is a numerical
challenge that could benefit from the various approaches developed
for the sign problem in finite density QCD.

\begin{acknowledgments}
The authors would like to thank Jac Verbaarschot for a discussion
on the RMT kernel.  This work used the Extreme Science and Engineering
Discovery Environment (XSEDE), which is supported by NSF under grant
number ACI-1548562.  Resources at Pittsburgh Supercomputing Center,
San Diego Supercomputer Center, LSU Center for Computation and
Technology and at University of Texas at Austin were used under the
XSEDE allocation TG-PHY170011. Some computations in this paper were
also performed on JLAB computing clusters under a USQCD type C
project.  R.N. acknowledges partial support by the NSF under grant
number PHY-1515446.  N.K. acknowledges support by the U.S. Department
of Energy under contract No. DE-SC0012704.
\end{acknowledgments}

\appendix
\section{Details of the lattice model}

We consider a symmetric periodic lattice with $L$ points in each
direction. Gauge fields on the lattice are denoted by $\theta_k({\bf
n})\in \mathbb{R} $ and they are associated with the links connecting
the sites ${\bf n}$ and ${\bf n}+\hat k$.  The fermions with charge
$q$ couple to the compact link variables
\be
U^q_k({\bf n}) = e^{iq \theta_k({\bf n})}.
\ee
It is essential for us to include a gauge action for the Abelian
field in order to be able to take the continuum limit at a fixed
physical volume $\ell^3$ of the torus, and have a continuum theory
free of parity anomaly.  For this, we use the non-compact plaquette
action given by
\be
S_g(\theta) = \frac{L}{\ell} \sum_{{\bf n}} \sum_{j < k =1}^3 \left(\Delta_j\theta_k({\bf n})-\Delta_k\theta_j({\bf n})\right)^2,
\label{ncaction}
\ee
where the lattice coupling is inversely proportional to the lattice
spacing $a=\frac{\ell}{L}$.  Monopoles are infinite energy
configurations in the continuum limit, and they are not part of the
path-integral.  Small values of $\ell$ correspond to the perturbative
limit and the non-perturbative aspects of the theory can be explored
by studying the asymptotic, $\ell\to\infty$, behavior  of physical
quantities after taking the continuum limit at a fixed $\ell$. The
non-compact plaquette action does not allow for non-zero net compact
flux, as seen by the fermion, over any cross-section of the torus.
In this case, the fermion charge $q$ is not required to be an
integer.

The regularized overlap-Dirac operator $\slashed{C}_o$ for a single
charge $q$ two-component fermion with a lattice mass $M\in [-1,1]$
is given by
\be 
 \slashed{C}_o(M,q\theta) = \frac{(1+M) +(1-M) V_{q\theta}}{2},
\ee
where $V_{q\theta}=\slashed{C}_W(\slashed{C}_W^\dagger \slashed{C}_W)^{-\frac{1}{2}}$
is a unitary operator constructed out of the
two-component gauge-link improved Sheikhoslami-Wohlert-Wilson-Dirac
operator $\slashed{C}_W$ with a negative
mass kept fixed in the range, $-M_W\in (0,2)$, as one takes
the continuum limit~\cite{Kikukawa:1997qh,Karthik:2016ppr}.  An
infra-red observable that we will consider later, is the spectrum
of the Hermitian operator $ iG^{-1}=i(1+V_{\theta})/(1-V_{\theta})$
near zero, where $G$ is the propagator of the $q=1$ massless overlap
fermion.

\section{Simulation details}

We used the Rational Hybrid Monte Carlo (RHMC)
algorithm~\cite{Kennedy:1998cu,Clark:2003na,Duane:1987de} for the
simulation.  In this method, the gauge fields are sampled using
essentially continuous molecular dynamics evolution, spoiled only
by the need to use a discrete time step in the numerical evolution
but rectified using accept-reject steps.  All gauge field configurations
along the evolution satisfy the importance sampling criterion as
per the positive definite measure and statistically independent
configurations are obtained by evolving for a finite time whose
value is decided by the autocorrelation time.

We improved the
overlap operator $\slashed{C}_{\rm o}$ by smoothening the gauge
fields $\theta$ that enter it by using one-level HYP smearing \ie, instead 
of $U^q_k(\mathbf{n})$, we used an improved 
link $V^q_k(\mathbf{n})=e^{iq \theta^s_k(\mathbf{n})}$ where $\theta^s_k(\mathbf{n})$ are 
HYP smeared~\cite{Karthik:2016ppr}. As explained in~\cite{Karthik:2016ppr}, this helps reducing any
non-zero monopole density at finite lattice spacing.  The overlap
operator is constructed using the Wilson-Dirac operator $\slashed{C}_W$
kernel, which we improved further using the Sheikloslami-Wohlert
coefficient $c_{\rm sw}=0.5$. We fixed the Wilson mass $M_W=1$ in
$\slashed{C}_W$ in all our simulations.

We included the fermion contribution $|\det \slashed{ C}_{\rm
o}|=|\det(\frac{1+V_\theta}{2})|$ by using its pseudo-fermion
representation:
\beq
|\det\slashed{ C}_{\rm o}| = \int [d\phi] e^{-\phi^{\dagger}\left[\slashed{ C}^\dagger_{\rm o}\slashed{ C}_{\rm o}\right]^{-1/2}\phi}.
\eeq{pseudof}
Using the standard procedure, we sampled $\phi$ from the above
distribution by sampling Gaussian distributed complex vectors $R$
through
\beq
\phi = \left[\slashed{C}^\dagger_{\rm o}\slashed{C}_{\rm o}\right]^{1/4}R.
\eeq{rvec}
We used the Zolotarev rational approximation $(r_k,p_k)$ for the above fourth-root:
\beq
\left[\slashed{ C}^\dagger_{\rm o}\slashed{ C}_{\rm o}\right]^{1/4} = r_0+\sum_{k=1}^{N_{\rm pole}}\frac{r_k}{\slashed{ C}^\dagger_{\rm o}\slashed{ C}_{\rm o}+p_k}.
\eeq{ratappx}
In the range of values for $\ell$ and $L$ we used, we found the
eigenvalues of $\slashed{C}^\dagger_{\rm o}\slashed{C}_{\rm o}$ to range at the most
from $10^{-6}$ to 1. We used the Remez algorithm to obtain the poles
$p_k$ and residues $r_k$ in the approximation with $N_{\rm pole}=20$,
such that the error in the approximation in the range $[10^{-7},1]$
is bounded by $10^{-8}$. We held these parameters fixed at all our
simulation points. 

With the usage of rational approximation, we used the standard
hybrid Monte Carlo (which now becomes the rational hybrid Monte
Carlo, RHMC) to sample gauge configurations --- we evolved the gauge
fields $\theta_i({\mathbf n})$ and the auxiliary momenta $\pi_i({\mathbf
n})$ conjugate to the gauge field through a fictitious molecular dynamics
time $\tau$ using the canonical equations of motion:
\beqa
\quad \frac{d}{d\tau}\theta_i({\mathbf n}) &=&\pi_i({\mathbf n});\cr
\frac{d}{d\tau}\pi_i({\mathbf n}) &=& -\frac{\partial}{\partial \theta_i({\mathbf n})}\left(\phi^{\dagger}\left[\slashed{ C}^\dagger_{\rm o}\slashed{ C}_{\rm o}\right]^{-1/2}\phi + S_g(\theta)\right).
\eeqa{canon}
Using another $N^\prime_{\rm pole}=20$ pole rational approximation 
$(r'_k,p'_k)$ for $\left[\slashed{ C}^\dagger_{\rm o}\slashed{ C}_{\rm o}\right]^{-1/2}$ 
(which approximates $x^{-1/2}$ within an error of $6\times 10^{-5}$ for $x\in[10^{-7},1]$), 
we get the fermionic contribution to the force $d\pi_i/d\tau$,
\beqa
\frac{\partial}{\partial \theta_i({\mathbf n})}\phi^{\dagger}\left[\slashed{C}^\dagger_{\rm o}\slashed{ C}_{\rm o}\right]^{-1/2}\phi &=& \sum_{k=1}^{N^\prime_{\rm pole}}r^\prime_k X^\dagger_k\frac{\partial \left(\slashed{ C}^\dagger_{\rm o}\slashed{ C}_{\rm o}\right)}{\partial\theta_i({\mathbf n})} X_k;\cr
X_k&=&(\slashed{ C}^\dagger_{\rm o}\slashed{ C}_{\rm o}+p^\prime_k)^{-1}\phi.
\eeqa{ratapp2}
The computation of the rest of the fermionic force calculation for each term
in the above sum is the same as the one given in~\cite{Karthik:2016ppr}.  

We evolved each trajectory for 1 unit of time $\tau$ ending with
an accept/reject step.  We tuned the step size $d\tau$ dynamically
during runtime such that the acceptance ratio was greater than 80\%.
After thermalization, we used only configurations separated by 5
trajectories for various measurements reported in this letter. At
each simulation point, we collected about 1000 such configurations
except in $L=24$ where the statistics is a bit smaller. The detailed
list of simulation points along with the parameters and measurements
are given in \tbn{data1} and \tbn{data2}.

\section{Measurement of eigenvalues, and the phase ${\cal A}$}
\bef
\vskip -1cm
\begin{center}
\includegraphics[scale=1.0]{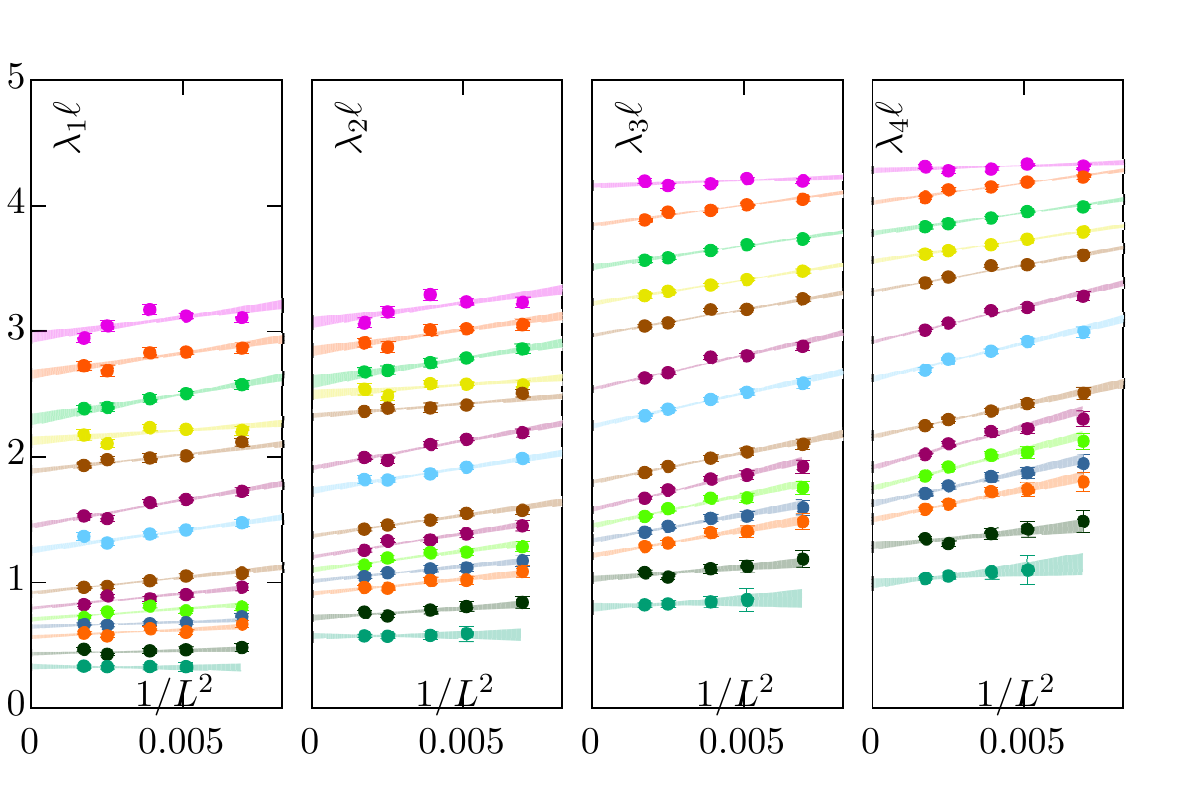}
\end{center}
\vskip -0.7cm
\caption{Continuum extrapolation of the eigenvalues $\lambda_i(\ell;L)\ell$ of $L|G^{-1}|$ for 
$i=1,2,3,4$ from the left to right respectively. In each panel, different colored symbols correspond to different 
fixed $\ell$; from the top to bottom of panels, $\ell=4,8,16,24,32,48.64,96,112,128,144,160,200$ and 250.
}
\eef{context}

For $L=6,8,10,12,14$ (and $L=16$ only for $\ell=200$), we constructed the unitary matrices $V_{\theta}$,
$V_{\theta/2}$ explicitly as $2L^3\times 2L^3$ matrices. Using
Lapack subroutines~\cite{laug}, we diagonalized these matrices to obtain their
eigenvalues $\exp(i\phi_j)$ and $\exp(i\Phi_j)$ respectively. Using
these eigenvalues, we constructed the phase ${\cal A}$ as
\beq
{\cal A}={\rm Im}\log\left(\prod_j^{2L^3} e^{-2i\Phi_j}\prod^{2L^3}_k (1+e^{i\phi_k})\right)\in (-\pi,\pi].
\eeq{phasecalc}
From the eigenvalues $e^{i\phi_j}$ of $V_{\theta}$, we also obtained
the eigenvalues $i\Lambda^{-1}_j=-i\cot(\phi_j/2)$ of the propagator
$G(\theta)$.  Since the computational cost of the brute force
eigenvalue computation is ${\cal O}(L^9)$, this was not a feasible
method for $L>14$. Instead, we used Ritz algorithm for $L=16,20,24$
to compute the four low-lying eigenvalues $\cos^2(\phi_j/2)$ of
$\slashed{C}^\dagger_{\rm o} \slashed{C}_{\rm o}$, from which we found the low-lying
eigenvalues $\Lambda_j$ of $|G^{-1}|$.

At finite $L$, these eigenvalues in lattice units are related to the 
continuum eigenvalues $\langle\lambda_j\ell\rangle$ through 
\beq
\left\langle\lambda_j\ell\right\rangle=2(M_W-M_t)\langle\Lambda_j\rangle L + \frac{c_1}{L^2}+\ldots,
\eeq{continuum}
where $M_W-M_t$ is the difference between the mass $M_W(=1)$ in the
Wilson-Dirac Kernel and the mass of the Wilson-Dirac fermion which
corresponds to the zero physical mass. We determined $M_t$ as the
Wilson mass where the smallest eigenvalue of $\slashed{C}^\dagger_W
\slashed{C}_W$ is minimized. One should note that $M_t\to 0$ in the
continuum limit and hence the usage of $2(M_W-M_t)$ instead of a
simpler $2M_W$ factor was only to improve the approach to the
continuum limit.  In the main text, the values of
$\langle\lambda_j(\ell;L)\rangle \ell$ are connected to the lattice
$\Lambda_j$ through $2(M_W-M_t)\langle\Lambda_j\rangle L$.  We have
tabulated these values of $\langle\lambda_j(\ell;L)\rangle \ell$
for $j=1,2,3,4$ for all the simulation points in \tbn{data1} and
\tbn{data2}.  In \fgn{context}, we show the $1/L^2$ extrapolation
of these improved low-lying eigenvalues to their continuum values
$\left\langle\lambda_j\ell\right\rangle$ using $L=12,14,16,20$ and
24.

\section{Random matrix theory}
The kernel for the random matrix theory appropriate for extracting
the condensate, if one exists, is given
by~\cite{Verbaarschot:1994ip,Damgaard:1997pw}
\be
K(x,y) = \frac{1}{2} \frac{\sqrt{|xy|}}{x-y} \left [ J_1(\pi x) J_0(\pi y) - J_0(\pi x) J_1(\pi y) \right].
\label{rmtkernel}
\ee
The procedure to extract the individual eigenvalue distribution is
standard~\cite{Nishigaki:2016nka}. We obtained the eigenvalues of
kernel numerically to a very good accuracy and used them to obtain the
individual eigenvalue distributions. The averages of the four lowest
distributions that appear in \eqn{sigdef} in the main text are $z_1=0.79787$,
$z_2=1.77186$, $z_3=2.763845$ and $z_4=3.76384$.

In \fgn{rmtdist} of the main text, we compare the distribution $P$ of the $i$-th scaled 
Dirac eigenvalue $\ell^2\Sigma_i\lambda_i$ and the distribition $P_{\rm RMT}$ of 
$i$-th eigenvalue from RMT. 
In order to quantify the approach of the Dirac eigenvalue distribution to the RMT 
distributions as $\ell\to\infty$, we consider
the sum of square deviations over the $N(=10)$ bins of the histogram,
\be
S=\sum_{z_{\rm bin}} \left(P(z_{\rm bin})-P_{\rm RMT}(z_{\rm bin})\right)^2.
\label{ssd}
\ee
In the left panel of \fgn{rmtmatch}, we show the square deviation $S$ for the 
four low-lying eigenvalue distributions
as a function of $\ell$, at fixed $L=20$. We find 
$S$ to decrease almost exponentially with $\ell$. On the right panel, we
show using the distribution of the smallest eigenvalue that 
the decrease in $S$ with $\ell$ remains robust as $L$ is increased. 
Thus, the agreement indeed 
gets better as one approaches the infinite volume limit for the first four eigenvalues. 
This is consistent with the
presence of a nonzero condensate in the infinite volume limit.

\bef
\begin{center}
\includegraphics[scale=1.25]{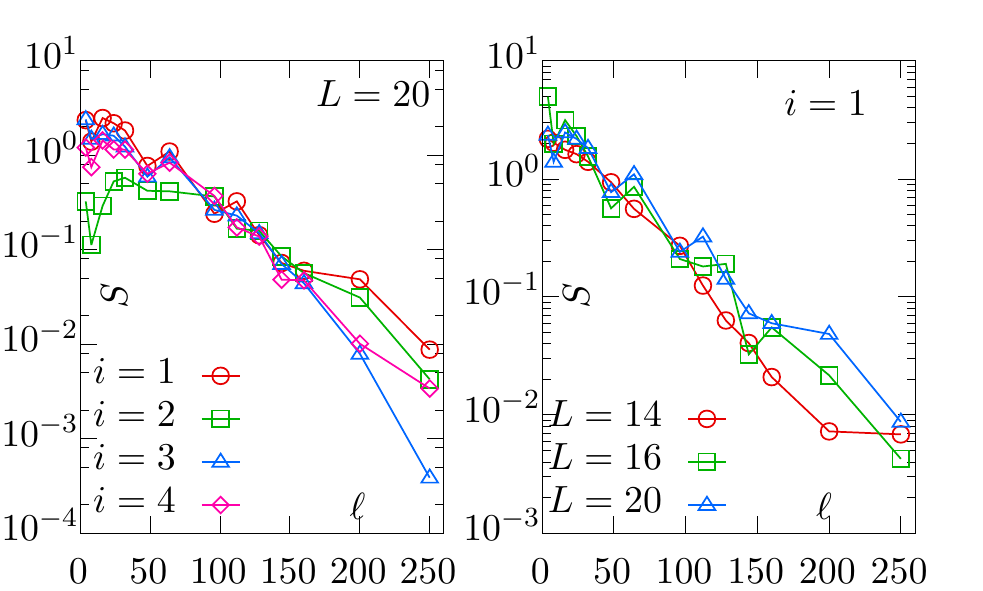}
\end{center}
\caption{
The square deviation $S$ of data from the RMT model as a function of the physical volume.
The $\ell$ dependences of $S$ for $i=1,2,3,4$ eigenvalue distributions at a fixed $L=20$ lattice 
is shown in the left panel. The lattice spacing effect on the $\ell$ dependence of $S$, for fixed
$i=1$, is shown in the right panel.
}
\eef{rmtmatch}

\bef
\begin{center}
\includegraphics[scale=1.25]{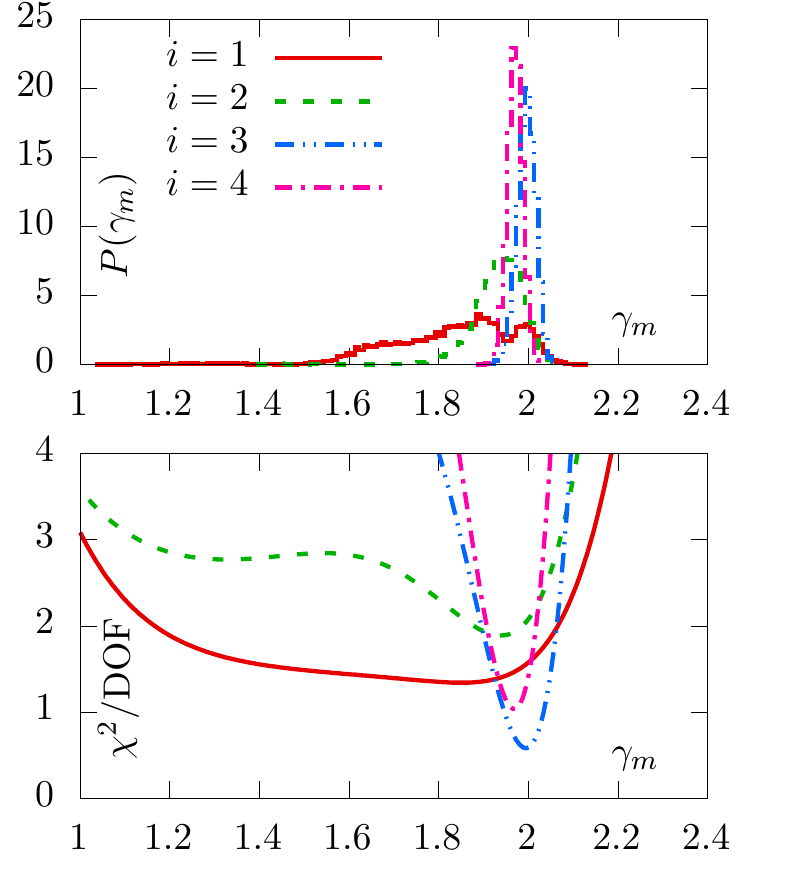}
\end{center}
\caption{
The $\chi^2/{\rm DOF}$, a measure of likelihood, of $\gamma_m$
corresponding to the asymptotic $\Sigma_i(\ell)\sim \ell^{-2+\gamma_m}$
scaling is shown in the bottom panel. The corresponding probability distribution 
of $\gamma_m$, as numerically constructed using
best-fit values of $\gamma_m$ in 5000 different 
bootstrap samples, is shown in the top panel. The distributions are peaked 
around $\gamma_m=2$ corresponding to the symmetry-breaking behavior.
}
\eef{gammam}

\section{Calculation of $\mathbf{J}^q$}
The essential simplification is
\be
2i q^2 \frac{ \delta {\cal A}_{\theta}}{\delta \theta_j({\bf n})} =
{\rm Tr} \left[ V^\dagger_{q\theta} \frac{ \delta V_{q\theta}}{\delta \theta_j({\bf n})}\right] = {\rm Im}{\rm Tr} \left[ \frac{1}{ C_W} \frac{\delta C_W }{\delta \theta_j({\bf n})}\right],
\ee
which is now in terms of the simpler Wilson-Dirac operator. One
should also note that the above expression means that the imaginary
part of the induced action from both the infinitely massive
Wilson-Dirac fermion as well as the corresponding overlap fermion
are the same. Thus, all our observations about the Wilson-Dirac
operator in~\cite{Karthik:2015sza} hold exactly for overlap fermion
as well.

\section{Likelihood of an infra-red fixed point}

In the analysis in the main text, we assumed $\Sigma(\ell=\infty)\ne
0$.  Instead, if the theory had an infra-red fixed point with a
mass anomalous dimension $0\le\gamma_m<2$, then $\Sigma_i(\ell)$
would become zero in the infinite volume with an $\ell^{-2+\gamma_m}$
asymptotic scaling.  As a check, we also tried to describe our
$\Sigma_i(\ell)$ data using a
$\Sigma_i(\ell)=k^\prime_1\ell^{-2+\gamma_m}\left(1+k^\prime_2\ell^{-1}+k^\prime_3\ell^{-2}\right)$
ansatz to include corrections to scaling, and we found that
$\chi^2/{\rm DOF}$ reaches a minimum, when $\gamma_m \approx 2$.
 We show
this behavior of $\chi^2$, along with the corresponding bootstrap
histogram of $\gamma_m$, in \fgn{gammam}. Even the lowest eigenvalue
that is affected by a soft edge (eigenvalues for a given configuration
is not symmetric around zero) shows evidence for a weak minimum for
$\gamma \approx 2$. Thus, we find consistent evidences pointing to
the presence of parity-breaking bilinear condensate.

\bibliography{biblio}

\begin{table*}
\centering
{\renewcommand{\arraystretch}{0.6}
\begin{tabular}{|c|l|l|l|l|l|l|l|}
\hline
\hline
$\ell$ & $L$ & $\lambda_1\ell$ & $\lambda_2\ell$ & $\lambda_3\ell$ & $\lambda_4\ell$ & $M_t$ & $N_{\rm conf}$ \\
\hline
4 & 6 & 3.296(38) & 3.412(39) & 4.544(16) & 4.649(16) & -0.0048(19) & 1008\\
 &8 & 3.133(57) & 3.255(59) & 4.327(27) & 4.443(26) & -0.0035(22) & 1008\\
 &10 & 3.164(47) & 3.287(49) & 4.260(17) & 4.376(17) & -0.0036(5) & 1008\\
 &12 & 3.112(42) & 3.233(43) & 4.198(21) & 4.312(21) & -0.0033(4) & 1008\\
 &14 & 3.121(21) & 3.236(22) & 4.217(9) & 4.333(9) & -0.0030(3) & 2064\\
 &16 & 3.173(41) & 3.293(42) & 4.175(15) & 4.294(15) & -0.0027(2) & 1008\\
 &20 & 3.043(46) & 3.155(47) & 4.161(18) & 4.280(19) & -0.0023(2) & 1008\\
 &24 & 2.946(38) & 3.066(39) & 4.198(20) & 4.315(19) & -0.0020(2) & 528\\
 \hline
8 &6 & 3.063(42) & 3.235(43) & 4.428(16) & 4.591(16) & -0.0019(38) & 1008\\
  &8 & 2.897(42) & 3.081(45) & 4.238(19) & 4.410(19) & -0.0035(22) & 1008\\
  &10 & 2.900(51) & 3.087(53) & 4.101(18) & 4.281(18) & -0.0041(12) & 1008\\
  &12 & 2.868(42) & 3.052(43) & 4.052(16) & 4.229(16) & -0.0040(6) & 1008\\
  &14 & 2.836(16) & 3.020(17) & 4.008(7) & 4.189(7) & -0.0041(5) & 2736\\
  &16 & 2.829(43) & 3.012(44) & 3.964(19) & 4.150(19) & -0.0040(5) & 1008\\
  &20 & 2.685(43) & 2.874(46) & 3.950(17) & 4.127(17) & -0.0035(3) & 1008\\
  &24 & 2.727(34) & 2.909(34) & 3.889(13) & 4.065(13) & -0.0032(3) & 696\\
\hline  
16 & 6 & 2.859(37) & 3.075(37) & 4.160(13) & 4.394(12) & 0.0131(111) & 1008\\
   &8 & 2.721(33) & 2.978(34) & 3.922(12) & 4.180(11) & 0.0075(45) & 1008\\
   &10 & 2.609(31) & 2.897(29) & 3.820(9) & 4.089(9) & 0.0024(29) & 1008\\
   &12 & 2.576(35) & 2.860(36) & 3.734(15) & 3.992(15) & -0.0000(19) & 1008\\
   &14 & 2.504(15) & 2.788(15) & 3.690(7) & 3.954(7) & -0.0015(13) & 3480\\
   &16 & 2.464(32) & 2.750(34) & 3.644(14) & 3.903(14) & -0.0023(9) & 1008\\
   &20 & 2.394(32) & 2.689(34) & 3.586(15) & 3.857(14) & -0.0034(6) & 1008\\
   &24 & 2.385(26) & 2.676(35) & 3.567(19) & 3.834(18) & -0.0038(5) & 816\\
\hline  
24 &6 & 2.718(23) & 2.919(23) & 3.968(15) & 4.253(14) & 0.0419(175) & 1008\\
   &8 & 2.534(38) & 2.807(34) & 3.713(12) & 4.018(10) & 0.0243(89) & 1008\\
   &10 & 2.390(38) & 2.696(35) & 3.555(12) & 3.872(12) & 0.0159(52) & 1008\\
   &12 & 2.208(35) & 2.571(36) & 3.480(18) & 3.791(19) & 0.0098(33) & 1008\\
   &14 & 2.219(11) & 2.579(11) & 3.413(8) & 3.735(8) & 0.0057(18) & 4560\\
   &16 & 2.234(29) & 2.583(28) & 3.370(14) & 3.691(12) & 0.0025(17) & 1008\\
   &20 & 2.107(35) & 2.485(33) & 3.319(14) & 3.643(13) & -0.0004(10) & 1008\\
   &24 & 2.175(15) & 2.539(14) & 3.285(8) & 3.616(8) & -0.0021(5) & 936\\
\hline
32   &6 & 2.631(35) & 2.798(30) & 3.773(16) & 4.090(12) & 0.0713(242) & 1008\\
&8 & 2.434(27) & 2.688(25) & 3.516(12) & 3.859(10) & 0.0507(122) & 1008\\
&10 & 2.183(41) & 2.502(40) & 3.357(12) & 3.721(12) & 0.0338(81) & 1008\\
&12 & 2.118(31) & 2.507(32) & 3.260(19) & 3.606(20) & 0.0223(46) & 1008\\
&14 & 2.008(11) & 2.413(12) & 3.175(12) & 3.532(13) & 0.0157(34) & 4368\\
&16 & 1.991(35) & 2.391(35) & 3.171(14) & 3.523(14) & 0.0109(23) & 1008\\
&20 & 1.977(25) & 2.389(25) & 3.068(13) & 3.432(11) & 0.0045(13) & 1008\\
&24 & 1.932(17) & 2.363(16) & 3.042(9) & 3.386(9) & 0.0012(8) & 984\\
\hline
48&6 & 2.508(30) & 2.596(23) & 3.421(11) & 3.795(11) & 0.1429(369) & 1008\\
&8 & 2.114(23) & 2.371(18) & 3.162(12) & 3.542(8) & 0.1026(219) & 1008\\
&10 & 1.945(26) & 2.280(25) & 2.988(11) & 3.383(10) & 0.0762(136) & 1008\\
&12 & 1.726(32) & 2.193(32) & 2.881(30) & 3.281(33) & 0.0571(90) & 1008\\
&14 & 1.662(14) & 2.140(15) & 2.805(18) & 3.191(21) & 0.0428(61) & 3840\\
&16 & 1.635(29) & 2.098(28) & 2.795(17) & 3.164(17) & 0.0330(42) & 1008\\
&20 & 1.509(24) & 1.973(25) & 2.671(14) & 3.067(13) & 0.0195(23) & 1008\\
&24 & 1.530(14) & 1.996(15) & 2.630(10) & 3.009(8) & 0.0121(14) & 984\\
\hline  
\end{tabular}
}
\caption{Simulation parameters $\ell$ and $L$,  eigenvalue data and statistics $N_{\rm conf}$.}
\label{tb:data1}
\end{table*}
\begin{table*}
\centering
{\renewcommand{\arraystretch}{0.6}
\begin{tabular}{|c|l|l|l|l|l|l|l|}
\hline
\hline
$\ell$ & $L$ & $\lambda_1\ell$ & $\lambda_2\ell$ & $\lambda_3\ell$ & $\lambda_4\ell$ & $M_t$ & $N_{\rm conf}$ \\
\hline
64&6 & 2.408(44) & 2.378(30) & 3.138(13) & 3.539(9) & 0.2124(526) & 1008\\
&8 & 1.921(37) & 2.140(26) & 2.836(11) & 3.270(11) & 0.1612(288) & 1008\\
&10 & 1.702(23) & 2.000(21) & 2.683(10) & 3.087(9) & 0.1250(178) & 1008\\
&12 & 1.476(28) & 1.988(33) & 2.585(38) & 2.991(43) & 0.0957(127) & 1008\\
&14 & 1.418(15) & 1.917(18) & 2.515(21) & 2.918(25) & 0.0748(77) & 3504\\
&16 & 1.385(23) & 1.864(22) & 2.458(18) & 2.842(19) & 0.0589(54) & 1008\\
&20 & 1.314(19) & 1.816(19) & 2.381(14) & 2.779(13) & 0.0385(31) & 1008\\
&24 & 1.366(34) & 1.820(33) & 2.326(19) & 2.692(20) & 0.0261(20) & 984\\
\hline
96&6 & 2.258(32) & 1.961(18) & 2.580(11) & 2.967(10) & 0.3621(689) & 1008\\
&8 & 1.700(28) & 1.779(19) & 2.389(10) & 2.790(8) & 0.2787(402) & 1008\\
&10 & 1.487(20) & 1.670(12) & 2.193(9) & 2.614(8) & 0.2249(261) & 1008\\
&12 & 1.072(27) & 1.573(34) & 2.099(41) & 2.506(48) & 0.1793(155) & 1008\\
&14 & 1.051(16) & 1.550(21) & 2.039(27) & 2.424(32) & 0.1446(111) & 2928\\
&16 & 1.014(20) & 1.497(21) & 1.990(23) & 2.365(25) & 0.1195(83) & 1008\\
&20 & 0.968(16) & 1.458(17) & 1.924(15) & 2.298(17) & 0.0846(55) & 1008\\
&24 & 0.962(10) & 1.424(13) & 1.876(12) & 2.249(12) & 0.0605(32) & 768\\
\hline
112&12 & 0.963(29) & 1.451(41) & 1.924(52) & 2.303(62) & 0.2187(206) & 1008\\
&14 & 0.903(19) & 1.385(27) & 1.853(36) & 2.227(43) & 0.1831(156) & 2640\\
&16 & 0.869(19) & 1.338(22) & 1.824(23) & 2.201(27) & 0.1500(96) & 1008\\
&20 & 0.893(14) & 1.333(16) & 1.737(16) & 2.103(18) & 0.1089(65) & 1008\\
&24 & 0.822(11) & 1.255(12) & 1.670(11) & 2.018(13) & 0.0811(37) & 672\\
\hline
128 &12 & 0.802(27) & 1.285(39) & 1.755(52) & 2.125(62) & 0.2631(214) & 1008\\
&14 & 0.775(20) & 1.241(29) & 1.673(39) & 2.036(47) & 0.2175(180) & 2280\\
&16 & 0.814(18) & 1.238(22) & 1.671(26) & 2.015(31) & 0.1800(118) & 1008\\
&20 & 0.764(14) & 1.195(17) & 1.587(18) & 1.920(20) & 0.1335(80) & 1008\\
&24 & 0.722(12) & 1.139(13) & 1.526(13) & 1.848(13) & 0.1010(49) & 600\\
\hline
144&12 & 0.727(30) & 1.173(45) & 1.594(60) & 1.946(72) & 0.3044(257) & 1008\\
&14 & 0.679(18) & 1.115(28) & 1.528(37) & 1.873(46) & 0.2524(181) & 2016\\
&16 & 0.670(17) & 1.108(22) & 1.511(27) & 1.844(33) & 0.2127(132) & 1008\\
&20 & 0.658(13) & 1.077(14) & 1.446(17) & 1.768(20) & 0.1587(84) & 1008\\
&24 & 0.666(11) & 1.046(13) & 1.400(13) & 1.708(15) & 0.1226(59) & 600\\
\hline
160&12 & 0.668(30) & 1.088(46) & 1.483(63) & 1.801(75) & 0.3389(275) & 1008\\
&14 & 0.603(21) & 1.016(34) & 1.406(46) & 1.739(57) & 0.2880(233) & 1870\\
&16 & 0.631(18) & 1.017(25) & 1.397(32) & 1.723(38) & 0.2450(163) & 1008\\
&20 & 0.575(12) & 0.954(15) & 1.314(18) & 1.624(21) & 0.1846(96) & 1008\\
&24 & 0.596(10) & 0.959(13) & 1.286(13) & 1.582(13) & 0.1437(56) & 480\\
\hline
200&12 & 0.481(29) & 0.842(49) & 1.186(68) & 1.486(85) & 0.4256(329) & 1008\\
&14 & 0.464(22) & 0.808(37) & 1.128(51) & 1.423(63) & 0.3733(276) & 1008\\
&16 & 0.455(16) & 0.780(25) & 1.109(35) & 1.389(43) & 0.3206(204) & 1008\\
&20 & 0.428(11) & 0.733(16) & 1.043(20) & 1.310(24) & 0.2491(126) & 983\\
&24 & 0.467(14) & 0.764(15) & 1.078(18) & 1.348(18) & 0.1958(90) & 359\\
\hline
250&14 & 0.329(35) & 0.590(62) & 0.860(90) & 1.098(115) & 0.4614(563) & 1008\\
&16 & 0.330(17) & 0.578(29) & 0.846(42) & 1.082(54) & 0.4153(285) & 1007\\
&20 & 0.329(11) & 0.571(17) & 0.828(23) & 1.051(29) & 0.3248(177) & 1006\\
&24 & 0.333(10) & 0.575(11) & 0.822(15) & 1.031(16) & 0.2616(94) & 240\\
\hline
\hline
\end{tabular}
}
\caption{Simulation parameters $\ell$ and $L$,  eigenvalue data and statistics $N_{\rm conf}$.}
\label{tb:data2}
\end{table*}
\end{document}